\PassOptionsToPackage{unicode}{hyperref}
\PassOptionsToPackage{hyphens}{url}
\PassOptionsToPackage{dvipsnames,svgnames,x11names}{xcolor}
\documentclass[
  12pt]{article}
\usepackage{amsthm}
\usepackage{amsfonts}
\usepackage{graphicx}
\usepackage{enumerate}
\usepackage{natbib}
\usepackage{url} 
\usepackage{color}
\usepackage{bm}
\usepackage{multirow}
\usepackage{pifont}
\usepackage{threeparttable}
\usepackage{booktabs}
\usepackage{caption}
\RequirePackage[colorlinks,citecolor=blue,urlcolor=blue,linkcolor=black]{hyperref}
\usepackage{abstract}
\usepackage{tikz}
\usepackage{enumitem}
\usepackage{amsmath,amssymb}
\usepackage{iftex}
\ifPDFTeX
  \usepackage[T1]{fontenc}
  \usepackage[utf8]{inputenc}
  \usepackage{textcomp} 
\else 
  \usepackage{unicode-math}
  \defaultfontfeatures{Scale=MatchLowercase}
  \defaultfontfeatures[\rmfamily]{Ligatures=TeX,Scale=1}
\fi
\usepackage{lmodern}
\ifPDFTeX\else  
\fi
\IfFileExists{upquote.sty}{\usepackage{upquote}}{}
\IfFileExists{microtype.sty}{
  \usepackage[]{microtype}
  \UseMicrotypeSet[protrusion]{basicmath} 
}{}
\makeatletter
\@ifundefined{KOMAClassName}{
  \IfFileExists{parskip.sty}{%
    \usepackage{parskip}
  }{
    \setlength{\parindent}{0pt}
    \setlength{\parskip}{6pt plus 2pt minus 1pt}}
}{
  \KOMAoptions{parskip=half}}
\makeatother
\usepackage{xcolor}
\makeatletter
\ifx\paragraph\undefined\else
  \let\oldparagraph\paragraph
  \renewcommand{\paragraph}{
    \@ifstar
      \xxxParagraphStar
      \xxxParagraphNoStar
  }
  \newcommand{\xxxParagraphStar}[1]{\oldparagraph*{#1}\mbox{}}
  \newcommand{\xxxParagraphNoStar}[1]{\oldparagraph{#1}\mbox{}}
\fi
\ifx\subparagraph\undefined\else
  \let\oldsubparagraph\subparagraph
  \renewcommand{\subparagraph}{
    \@ifstar
      \xxxSubParagraphStar
      \xxxSubParagraphNoStar
  }
  \newcommand{\xxxSubParagraphStar}[1]{\oldsubparagraph*{#1}\mbox{}}
  \newcommand{\xxxSubParagraphNoStar}[1]{\oldsubparagraph{#1}\mbox{}}
\fi
\makeatother

\usepackage{longtable,booktabs,array}
\usepackage{calc} 
\usepackage{etoolbox}
\makeatletter
\patchcmd\longtable{\par}{\if@noskipsec\mbox{}\fi\par}{}{}
\makeatother
\IfFileExists{footnotehyper.sty}{\usepackage{footnotehyper}}{\usepackage{footnote}}
\makesavenoteenv{longtable}
\usepackage{graphicx}
\makeatletter
\def\maxwidth{\ifdim\Gin@nat@width>\linewidth\linewidth\else\Gin@nat@width\fi}
\def\maxheight{\ifdim\Gin@nat@height>\textheight\textheight\else\Gin@nat@height\fi}
\makeatother
\setkeys{Gin}{width=\maxwidth,height=\maxheight,keepaspectratio}
\makeatletter
\def\fps@figure{htbp}
\makeatother

\makeatletter
\@ifpackageloaded{caption}{}{\usepackage{caption}}
\AtBeginDocument{%
\ifdefined\contentsname
  \renewcommand*\contentsname{Table of contents}
\else
  \newcommand\contentsname{Table of contents}
\fi
\ifdefined\listfigurename
  \renewcommand*\listfigurename{List of Figures}
\else
  \newcommand\listfigurename{List of Figures}
\fi
\ifdefined\listtablename
  \renewcommand*\listtablename{List of Tables}
\else
  \newcommand\listtablename{List of Tables}
\fi
\ifdefined\figurename
  \renewcommand*\figurename{Figure}
\else
  \newcommand\figurename{Figure}
\fi
\ifdefined\tablename
  \renewcommand*\tablename{Table}
\else
  \newcommand\tablename{Table}
\fi
}
\@ifpackageloaded{float}{}{\usepackage{float}}
\floatstyle{ruled}
\@ifundefined{c@chapter}{\newfloat{codelisting}{h}{lop}}{\newfloat{codelisting}{h}{lop}[chapter]}
\floatname{codelisting}{Listing}

\makeatother
\makeatletter
\@ifpackageloaded{caption}{}{\usepackage{caption}}
\@ifpackageloaded{subcaption}{}{\usepackage{subcaption}}
\makeatother
\DeclareMathOperator*{\argmin}{arg\,min}

\newcommand{\blind}{1}

\newtheorem{theorem}{Theorem}
\newtheorem{lemma}{Lemma}

\newtheorem{assumption}{Assumption}

\def\spacingset#1{\renewcommand{\baselinestretch}%
{#1}\small\normalsize} \spacingset{1}

\ifLuaTeX
  \usepackage{selnolig}  
\fi
\usepackage[]{natbib}
\usepackage{bookmark}

\IfFileExists{xurl.sty}{\usepackage{xurl}}{} 
\hypersetup{
  pdftitle={Title},
  pdfauthor={Author 1; Author 2},
  pdfkeywords={3 to 6 keywords, that do not appear in the title},
  colorlinks=true,
  linkcolor={blue},
  filecolor={Maroon},
  citecolor={Blue},
  urlcolor={Blue},
  pdfcreator={LaTeX via pandoc}}

\definecolor{myblue}{RGB}{33,59,251} 
\definecolor{lightgreen}{RGB}{146,208,80} 
\definecolor{ZurichRed}{rgb}{1, 0, 0} 
\definecolor{ZurichGreen}{rgb}{.196,.804,.196} 
\definecolor{ZurichOrange}{rgb}{1,.648,0} 
	\definecolor{gold}{rgb}{0.83, 0.69, 0.22}

\newcommand{\ind}{\mbox{$\perp \kern-5.5pt \perp$}}

\newcommand{\cmark}{\ding{51}}%
\newcommand{\xmark}{\ding{55}}%

\usetikzlibrary{arrows,shapes.arrows,shapes.geometric,
	shapes.multipart,backgrounds,decorations.pathmorphing,positioning,fit,automata}
\tikzset{
	>=stealth',
	truetwo/.style={
		rectangle,
		draw=black, very thick,
		text width=8.5em,
		minimum height=2em,
		text centered,
		fill=blue!20, opacity = 1.0},
	trueone/.style={
		rectangle,
		draw=black, very thick,
		text width=8.5em,
		minimum height=2em,
		text centered,
		fill=gray!50, opacity = 1.0},
	truethree/.style={
		rectangle,
		draw=black, very thick,
		text width=8.5em,
		minimum height=2em,
		text centered,
		fill=orange, opacity = 1.0},
	punkt/.style={
		rectangle,
		rounded corners,
		draw=black, very thick,
		text width=6.5em,
		minimum height=2em,
		text centered},
	est/.style={
		rectangle,
		draw=black, very thick,
		text centered},
	mytext/.style={
			rectangle,
			draw=white, very thick,
			text width=6.5em,
			minimum height=2em,
			text centered},
	shade/.style={
		rectangle,
		draw=black, very thick, fill=gray!50,
		text centered},
	weight/.style={
		circle,
		draw=black, very thick,
		text width=6.5em,
		minimum height=2em,
		text centered},
	pil/.style={
		->,
		thick,
		shorten <=2pt,
		shorten >=2pt,},
	double/.style={
		<->,
		thick,
		shorten <=2pt,
		shorten >=2pt,},
	dash/.style={
		dashed,
		thick,
		shorten <=2pt,
		shorten >=2pt,},
	dashdouble/.style={
		<->,
		dashed,
		thick,
		shorten <=2pt,
		shorten >=2pt,},
LargeBlock/.style={rectangle, draw, fill=blue!20, text width=10em, text badly centered, rounded corners},
decision/.style = {diamond, draw, fill=red!20, text badly centered,text width=1.2cm},
block/.style = {rectangle, draw, fill=blue!20, text width=5em, text badly centered, rounded corners, minimum height=1em},
ImgBlock/.style = {rectangle, draw},
line/.style = {draw, -latex'},
cloud/.style = {draw, ellipse,fill=red!20, minimum height=1em},
EmptyAnchor/.style = {circle,minimum height=1em}

}

\usepackage{pdfrender}

\newcommand{\anon}{1}

\begin{document}

\def\spacingset#1{\renewcommand{\baselinestretch}%
{#1}\small\normalsize} \spacingset{1}

\if1\anon
{
	 \begin{center} 
	\spacingset{1.5} 	{\LARGE\bf  Constructive Instrumental Variable Identification and Inference with Many Weak Interaction Moments} \\ \bigskip \bigskip
		\spacingset{1} 
		{\large
Di Zhang$^1$, Minhao Yao$^2$, Zhonghua Liu$^3$\footnote{Co-corresponding authors: Zhonghua Liu, Department of  Biostatistics, Columbia University, 722 West 168th Street, New York, NY, 10032, USA;  E-mail: zl2509@cumc.columbia.edu. Baoluo Sun,  Department of Statistics and Data Science, National University of Singapore, Singapore; E-mail: stasb@nus.edu.sg}, and Baoluo Sun$^{1\ast}$\\ \bigskip
\normalsize
$^1$ Department of Statistics and Data Science, National University of Singapore\\ 
\smallskip
$^2$ Centre for Biomedical Data Science, Duke-NUS Medical School, National University of Singapore \\ \smallskip
$^3$ Department of  Biostatistics, Columbia University
}

	{\it 
}
	\end{center}
} \fi

\if0\blind
{
  \bigskip
  \bigskip
  \bigskip
  \begin{center}\spacingset{1.5} 
    {\Large\bf Constructive Instrumental Variable Identification and Inference with Many Weak Interaction Moments}
\end{center}
  \medskip
} \fi

\bigskip

\begin{abstract}
Instrumental variable methods are widely used for causal inference, but identification becomes especially challenging when instruments are weak and potentially invalid. These challenges are particularly pronounced in Mendelian randomization, where genetic variants serve as instruments and violations of exclusion restriction or independence assumptions are common. We propose MAGIC, a constructive and assumption-lean framework that achieves identification even when all candidate instruments may be invalid. The method exploits pairwise and higher-order interactions among mutually independent instruments to construct moment conditions orthogonal to both unmeasured confounding and direct effects under a linear structural model. The resulting estimation problem involves many potentially weak interaction moments with unknown nuisance parameters. We develop a semiparametric generalized method of moments estimator and introduce a global Neyman orthogonality condition to ensure robustness of both the moment function and its derivative to nuisance estimation under many weak moment asymptotics. We establish consistency and asymptotic normality when the number of moments diverges with sample size and characterize the semiparametric efficiency bound under fixed dimension. Simulations and an application to UK Biobank data illustrate the method.

\end{abstract}

\noindent%
\vfill

\newpage
\spacingset{1.8} 

\section{Introduction}\label{sec: intro}

Understanding causal relationships in observational studies is a central challenge across scientific disciplines. Instrumental variable (IV) methods provide a general framework for addressing this challenge when exogenous variation is available \citep{angrist1994identification,angrist1995two,hernan2006instruments,kang2024identification}. Mendelian randomization (MR) is a prominent example, leveraging genetic variants, typically single nucleotide polymorphisms (SNPs), as instruments to estimate the causal effect of an exposure on an outcome \citep{Davey-Smith:2003aa,didelez2007mendelian,lawlor:2008aa,sanderson2022mendelian,Yao2026}. MR studies may be viewed as natural experiments that exploit the random allocation of genetic variants at conception to generate biologically grounded exogenous variation in the exposure. This principle forms the conceptual basis of MR.

Under the conventional MR framework \citep{didelez2007mendelian}, valid IVs must satisfy three core assumptions: (A1) relevance, (A2) exclusion restriction, and (A3) independence. In practice, violations of these assumptions are common and can severely bias causal estimates \citep{vanderweele2014methodological}. Weak instrument bias, corresponding to near violation of A1, pleiotropy, corresponding to violation of A2, and population stratification or linkage disequilibrium, corresponding to violation of A3, are pervasive in analyses of complex traits \citep{Yao2026}. These issues reflect broader challenges in IV models when instruments are weak or invalid.

To accommodate violations of IV assumptions, a substantial literature has developed under the additive linear constant effects (ALICE) model of \citet{holland1988causal}, extended to allow multiple invalid instruments \citep{Kang:2016aa}. Existing approaches under the ALICE framework fall largely into two categories. The first class of methods relies on majority or plurality rules, or assumes that a pre-specified number of instruments are valid, thereby imposing restrictions on how many instruments satisfy the validity conditions \citep{han2008detecting,Kang:2016aa,Windmeijer:2019aa,Guo:2018aa,windmeijer2021confidence,guo2023causal, sun2023semiparametric,10.1093/jrsssb/qkae025}. The second allows all instruments to be invalid but relies on structural  assumptions, such as the independence between instrument strength and direct effects (InSIDE) or balanced pleiotropy, typically formulated within random-effects models \citep{kolesar2015identification,zhao2018statistical}. See \citet{kang2024identification} for a recent review and references therein. 

Despite this progress, two fundamental statistical challenges remain. First, many identification strategies rely on structural assumptions about instrument validity that are difficult to verify empirically. Second, modern genetic studies frequently involve many potentially weak signals, leading to settings in which the number of moment conditions grows with sample size. Classical generalized method of moments (GMM) theory \citep{Hansen:1982aa,Newey:2009aa} and  Neyman orthogonality  \citep{neyman1,neyman2,Chernozhukov2018ddml} are not directly tailored to this regime, particularly when nuisance parameters must be estimated and the dimension of the moment vector diverges. In such settings, small biases from nuisance estimation can accumulate across many weak moments, potentially invalidating standard asymptotic approximations.

Our approach is motivated by a simple observation: while individual instruments may violate exclusion or independence assumptions, interactions among mutually independent instruments behave differently. In particular, higher-order interactions can be constructed to be orthogonal to bias-inducing components of the outcome model, even when all main-effect instruments are invalid. This observation suggests that identification can be achieved not by selecting valid instruments, but by constructing valid moment conditions from interactions.

We propose a new framework, MAGIC, that addresses both challenges. MAGIC provides a constructive and assumption-lean approach to identification and enables valid semiparametric inference under the ALICE model. A key distinction from prior work is that the identifying condition, instrument independence, is empirically testable. We refer to this as \emph{constructive identification} to emphasize that the identifying moment conditions arise from empirical design, for example through linkage disequilibrium (LD) clumping to obtain approximately independent variants \citep{purcell2007plink}, combined with interaction construction. We describe the framework as \emph{assumption-lean} because identification neither requires any instruments to be valid nor imposes restrictions on the number or distribution of invalid instruments. Our contributions are threefold.

First, we propose an interaction-based identification strategy that exploits pairwise and higher-order products of independent instruments. These interaction terms are orthogonal to bias-inducing direct effects and unmeasured confounding under the ALICE model, enabling identification even when all candidate instruments are invalid. Identification therefore follows from empirically verifiable independence and interaction construction. Evidence that genetic interactions, also known as epistasis \citep{zuk2012mystery,wei2014detecting}, are common in complex traits  further supports the practical plausibility of interaction-based identification. This shifts the focus from selecting valid instruments to constructing valid moment conditions.

Second, we characterize the semiparametric efficiency bound for estimation of the causal parameter under the interaction-based moment restriction when the number of interaction moments is fixed, and show that the proposed estimator attains this bound. 

Third, to enable valid inference under the proposed interaction-based identification strategy, we develop asymptotic theory for estimation and inference in the presence of many weak conditions with estimated nuisance parameters. We introduce a global Neyman orthogonality condition that extends classical (local) Neyman orthogonality \citep{neyman1,neyman2,Chernozhukov2018ddml} to settings where the number of moment conditions diverges with the sample size. In this regime, nuisance estimation error enters both the empirical moment condition and its derivative, and the resulting bias accumulates as the dimension of the moment vector increases. We show that global Neyman orthogonality eliminates this accumulationat first order and ensures valid inference. Classical GMM theory \citep{Hansen:1982aa,Newey:2009aa} does not address this setting with estimated nuisance parameters, and related work such as \citet{ye2024genius} relies on particular moment constructions. Around the time this work was completed, \citet{wang2025gmm} independently studied GMM with many weak conditions and nuisance parameters in a general framework. Our analysis centers on constructive identification via IV interactions and clarifies the role of global Neyman orthogonality in enabling inference under this many weak interaction asymptotic regime, based on a novel Robinson-style \citep{robinson1988root} moment function carefully constructed to satisfy this global orthogonality condition.

The remainder of the paper is organized as follows. Section~\ref{sec:prelim} introduces the setup of the model and the constructive identification strategy. Section~\ref{sec:semi} develops the semiparametric efficiency theory. Section~\ref{sec:challenges} presents the MAGIC estimator and corresponding inference procedures under many weak interaction moments, with theoretical results established in Section~\ref{sec:the}. Simulation studies are reported in Section~\ref{sec:sim}, followed by an application to UK Biobank data in Section~\ref{sec:app}. Section~\ref{sec:discussion} concludes.

\section{Constructive Identification}

\label{sec:prelim}

{
\subsection{Data Structure and Notation}

To describe the model, let $({O}_1,\ldots,{O}_n)$ denote independent and identically distributed observations of the random vector ${O} = (Y, D, Z)$, where $Y$ is the outcome of interest, $D$ is an exposure or treatment variable, and $Z = (Z_1,\ldots,Z_p)$ is a vector of $p \geq 2$ candidate instrumental variables (IVs). The support of $O$ is denoted by $\mathcal{O} = \mathcal{Y} \times \mathcal{D} \times \mathcal{Z}$. To ease exposition, we focus on the canonical case where the IVs are binary, i.e., $\mathcal{Z} = \{0,1\}^p$, although our proposed methodology applies more broadly to discrete or continuous IVs. Let $\mu^* = (\mu_1^*, \ldots, \mu_p^*) = \mathbb{E}(Z)$ denote the vector of population means.

 The following notation will also be used throughout the paper. For a matrix or vector $A$, its transpose is denoted as $A^\top$, and $A^{\otimes 2}=AA^\top$. Let $|S|$ denote the cardinality of a set $S$. To describe the interaction of candidate IVs,  for any $1\leq k\leq p$, let $S_k=\{A\subseteq \{1,2,...,p\}: |A|=k\}$ denote the collection of all distinct size-$k$ subsets of $\{1,2,...,p\}$, and let $\bar{Z}_{k,\mu}$ denote the vector of all distinct, demeaned  $k$-th order interactions of the candidate IVs, composed of elements from the set $\{\prod_{j\in x}(Z_j-\mu_j): x \in S_k\}$ in some specific order. We also let $\bar{Z}_{k}=\bar{Z}_{k,0}$ denote the vector of non-demeaned $k$-th order interactions. For example, if $p=3$, then $\bar{Z}_{2,\mu}=\{(Z_1-\mu_1)(Z_2-\mu_2),(Z_1-\mu_1)(Z_3-\mu_3),(Z_2-\mu_2)(Z_3-\mu_3)\}^\top$ and $\bar{Z}_{2}=(Z_1 Z_2, Z_1 Z_3, Z_2 Z_3)^\top$. In addition, let $\mathbb{E}_n f=n^{-1}\sum^n_{i=1} f_i=n^{-1}\sum^n_{i=1} f(O_i)$ denote the empirical mean for an arbitrary  measurable real-valued function $f$, and we use $\partial f(\theta^{\ast})/\partial \theta$ to abbreviate the partial derivatives $\partial f(\theta)/\partial \theta\mid_{\theta=\theta^{\ast}}$.

\subsection{The ALICE Model with Possibly Invalid Instruments}
Let \( Y(d, z) \) denote the potential outcome that would be observed if the exposure and instrumental variables (IVs) were set to levels \( d \in \mathcal{D} \) and \( z \in \mathcal{Z} \), respectively. For two exposure levels \( d', d \in \mathcal{D} \) and two IV configurations \( z', z \in \mathcal{Z} \), we adopt the following structural model, known as the ALICE model (Additive Linear, Independent, Constant Effects) \citep{holland1988causal}, which has been widely used to accommodate invalid IVs \citep{small2007sensitivity, Kang:2016aa, Guo:2018aa, Windmeijer:2019aa, kang2020two, windmeijer2021confidence, guo2023causal}.

\begin{assumption}[ALICE Model]
\label{assp:alice}
\[
Y(d', z') - Y(d, z) = (d' - d)\beta^* + (z' - z)\zeta^*, 
\qquad 
\mathbb{E}\{Y(0, 0) \mid Z\} = Z \psi^*.
\]
\end{assumption}

The scalar parameter \( \beta^* \in \mathbb{R} \) represents the causal effect of the exposure on the outcome. The vector \( \zeta^* \in \mathbb{R}^p \) captures direct effects of the IVs on the outcome, with \( \zeta^*_j \) denoting the direct effect of the \( j \)-th IV. Assumption \ref{assp:alice} also allows the baseline potential outcome \( Y(0,0) \) to depend linearly on \( Z \), thereby accommodating associations between the IVs and unmeasured confounders through the vector \( \psi^* \in \mathbb{R}^p \). Here, \( \psi^*_j \) reflects the confounding bias associated with the \( j \)-th IV.

Under Assumption \ref{assp:alice}, the potential outcome can be written as
\[
Y(d,z) = Y(0,0) + d\beta^* + z\zeta^*.
\]
Evaluating at the observed exposure and IV values \( (D, Z) \) and invoking causal consistency \( Y = Y(D,Z) \), we obtain
\[
Y = Y(0,0) + D\beta^* + Z\zeta^*.
\]
Using the decomposition
\[
Y(0,0) = Z\psi^* + \epsilon,
\quad 
\text{where } 
\epsilon := Y(0,0) - \mathbb{E}\{Y(0,0)\mid Z\},
\]
we have \( \mathbb{E}(\epsilon \mid Z) = 0 \) by construction. Substituting this expression yields the observed data model
\begin{equation}
\label{lout}
Y = D\beta^* + Z(\zeta^* + \psi^*) + \epsilon, 
\qquad 
\mathbb{E}(\epsilon \mid Z) = 0.
\end{equation}

Letting \( \pi^* = \zeta^* + \psi^* \), the vector \( \pi^* \) summarizes violations of the exclusion restriction (A2) and independence (A3) assumptions. The \( j \)-th IV is therefore valid if \( \pi^*_j = 0 \) and invalid otherwise.

\subsection{Identification via Constructed  IV Interactions}

Identification of the causal effect $\beta^{*}$ is commonly justified by assumptions such as the existence of valid instruments, majority or plurality validity, or balanced pleiotropy. These assumptions are generally unverifiable and may fail in complex trait settings \citep{Yao2026}, potentially leading to biased causal inference.

We take a different approach. Rather than restricting the number or structure of invalid instruments, we construct interaction-based moment conditions using pairwise and higher-order products of instrumental variables. When the candidate single nucleotide polymorphisms (SNPs) are mutually independent, these interaction terms generate orthogonality conditions that remain valid even if all main-effect instruments are invalid. 

\begin{assumption}[Independent IVs]
\label{assp:indp}
The $p$ candidate instrumental variables  are mutually independent.
\end{assumption}

Assumption \ref{assp:indp} requires that the selected SNPs are not in LD in MR studies. In contrast to structural assumptions on instrument validity, this condition can be empirically approximated through standard LD-clumping procedures \citep{purcell2007plink, zhao2018statistical, hemani2018mr}. 

Under Assumption \ref{assp:indp}, interaction terms  are orthogonal to the linear violations captured by the ALICE model, enabling identification provided that at least one constructed interaction is associated with the exposure. Genetic evidence suggests that epistatic effects are common and may contribute meaningfully to trait variation \citep{zuk2012mystery, wei2014detecting}, supporting the practical plausibility of interaction-based identification.

\subsubsection*{Example: Identification via a Constructed Pairwise IV Interaction}

To illustrate the identification strategy, consider a simple example involving a pairwise interaction between two candidate IVs. Suppose $(Z_1, Z_2)$ are two mutually independent IVs with means $\mu_1^*$ and $\mu_2^*$. Consider the moment condition
\begin{equation}
\label{eq:zero}
\mathbb{E}\{(Z_1 - \mu_1^*)(Z_2 - \mu_2^*)(Y - D\beta)\} = 0.
\end{equation}

We show that this condition holds at the true causal effect $\beta = \beta^*$ under Assumptions~\ref{assp:alice} and~\ref{assp:indp}. Under the ALICE model,
\[
\mathbb{E}(Y - D\beta^* \mid Z) = Z \pi^*,
\]
where $\pi^* = (\pi_1^*, \dots, \pi_p^*)^\top$ summarizes violations of the classical IV assumptions. Taking iterated expectations,
\begin{align*}
\mathbb{E}\left[(Z_1 - \mu_1^*)(Z_2 - \mu_2^*)(Y - D\beta^*)\right]
&= \mathbb{E}\left[(Z_1 - \mu_1^*)(Z_2 - \mu_2^*) \, \mathbb{E}(Y - D\beta^* \mid Z)\right] \\
&= \sum_{j=1}^p \pi_j^* \, 
\mathbb{E}\left[(Z_1 - \mu_1^*)(Z_2 - \mu_2^*) Z_j\right].
\end{align*}

By mutual independence (Assumption~\ref{assp:indp}), the centered product 
$(Z_1 - \mu_1^*)(Z_2 - \mu_2^*)$ is orthogonal to each $Z_j$, so each expectation above equals zero. Therefore, the moment condition \eqref{eq:zero} holds regardless of the values of $\pi^*$. In particular, it remains valid even if all candidate IVs are invalid.

Moreover, if the interaction term is relevant for the exposure, i.e.,
\[
\mathbb{E}[(Z_1 - \mu_1^*)(Z_2 - \mu_2^*)D] \neq 0,
\]
then equation \eqref{eq:zero} uniquely identifies
\[
\beta^* = 
\frac{\mathbb{E}[(Z_1 - \mu_1^*)(Z_2 - \mu_2^*)Y]}
{\mathbb{E}[(Z_1 - \mu_1^*)(Z_2 - \mu_2^*)D]}.
\]

This argument does not depend on the specific pair $(Z_1, Z_2)$. More generally, one may construct all distinct demeaned interaction terms up to order $q$, for $2 \leq q \leq p$. The total number of such interactions is
$
r(p,q) = \sum_{k=2}^q {p \choose k}.
$
Each interaction generates a valid moment condition under 
Assumptions~\ref{assp:alice} and \ref{assp:indp}. 
Collecting all demeaned interaction terms up to order $q$, 
define the $r(p,q)$-dimensional moment function
\[
m(O;\beta,\mu^*)
=
(\bar{Z}^\top_{2,\mu^{\ast}},\dots,\bar{Z}^\top_{q,\mu^{\ast}})^\top
(Y-D\beta).
\]
Identification of causal effect $\beta^*$ requires that at least one of these constructed 
interactions is relevant for the exposure. We formalize this condition below.

\begin{assumption}[Interaction Relevance]
\label{assp:relevance}
There exists at least one interaction term among 
$\{\bar{Z}_{2,\mu^*}, \dots, \bar{Z}_{q,\mu^*}\}$ 
such that the $r$-dimensional vector
\begin{align}
\label{eq:M}
M=\mathbb{E}\{\partial m(O;\beta^{\ast},\mu^{\ast})/\partial\beta\}=-\mathbb{E}\{(\bar{Z}^\top _{2,\mu^{\ast}},...,\bar{Z}^\top _{q,\mu^{\ast}})^\top D\}
\neq 0.
\end{align}
\end{assumption}

Assumption~\ref{assp:relevance} requires that at least one constructed 
interaction term is correlated with the exposure. 
Under this condition, the interaction-based moment restrictions 
provide sufficient variation to identify the causal effect.

\begin{theorem}[Constructive Identification]
\label{thm:moment}
Suppose Assumptions~\ref{assp:alice}, \ref{assp:indp}, and 
\ref{assp:relevance} hold. Then the true parameter value 
$\beta=\beta^{\ast}$ is the unique solution to the population 
moment condition
\begin{equation}
  \label{eq:gmm}
\mathbb{E}\{m(O;\beta,\mu^{\ast})\}=0.  
\end{equation}
\end{theorem}

\noindent
Theorem~\ref{thm:moment} establishes that identification follows from 
interaction-based moment construction together with a relevance condition 
on the constructed interactions. In particular, the causal parameter is 
identified without imposing restrictions on the validity of the putative IVs, but from variation generated by the 
interaction terms themselves.

\section{Semiparametric Efficiency Theory}
\label{sec:semi}
In this section, we characterize the semiparametric efficiency bound for estimation of the causal parameter $\beta^*$ defined by the interaction-based moment restriction in Theorem~\ref{thm:moment}, under the classical setting where the dimension $r$ of the moment vector is fixed. This fixed-dimensional analysis serves two purposes. First, it clarifies the efficient influence function associated with the constructive identification strategy. Second, it provides a benchmark for comparison with the many weak interaction asymptotic regime studied in Sections~4 and~5, where the number of moments diverges with the sample size. 

When the number of moment conditions $r \geq 2$, the model is overidentified and estimation proceeds within the generalized method of moments (GMM) framework \citep{Hansen:1982aa}. We begin with the benchmark case in which the analyst has access to the true population means of the candidate IVs, denoted by $\mu^* = (\mu_1^*,\dots,\mu_p^*)^\top$. In this setting, the GMM estimator is defined as
\[
\tilde{\beta}_{\mathrm{GMM}} 
= \argmin_{\beta \in \mathcal{B}} 
\frac{1}{2}\hat{m}(\beta)^\top \hat{W}\hat{m}(\beta),\quad \hat{m}(\beta) = \mathbb{E}_n\{m(O; \beta,\mu^*)\},
\]
where $\mathcal{B} \subset \mathbb{R}$ is a compact parameter space and $\hat{W}$ is a symmetric positive semi-definite $r \times r$ weighting matrix. In practice, the population means in the empirical moment vector are unknown and must be estimated; plugging in of these estimated means can affect the 
first-order asymptotic behavior of GMM estimators based on the moment 
restriction~\eqref{eq:gmm} with fixed $r$. Specifically, we can show that regular and asymptotically linear  GMM estimators 
$\tilde\beta$ with estimated means described in the supplementary material admit the expansion
\[
\sqrt{n}(\tilde{\beta}-\beta^*)
=
(\theta^\top M)^{-1}
\sqrt{n}\,
\mathbb{E}_n\!\left\{
\theta^\top 
\bar m(O;\beta^*,\mu^*,\pi^*)
\right\}
+ o_p(1),
\]
where
\[
\bar m(O;\beta^*,\mu^*,\pi^*)
=
(\bar Z_{2,\mu^*}^\top,\dots,\bar Z_{q,\mu^*}^\top)^\top
(Y-D\beta^*-Z\pi^*),
\]
and $\theta\in\mathbb{R}^r$ satisfies $\theta^\top M\neq0$ where $M$ is defined in equation (\ref{eq:M}). 
\begin{theorem}[Semiparametric Efficiency Bound]
\label{thm:eif}
Under Assumptions~\ref{assp:alice}--~\ref{assp:relevance} with fixed  $r$, 
the efficient influence function for $\beta^*$ is indexed by $\theta_{\text{opt}} = \Omega^{-1} M$,
where
\[
\Omega
=
\mathbb{E}\{\bar m(O;\beta^*,\mu^*,\pi^*)^{\otimes 2}\}.
\]
The corresponding semiparametric efficiency bound is
$
(M^\top \Omega^{-1} M)^{-1}.
$
\end{theorem}

The bound can be attained by GMM estimators with optimal weighting matrix 
\citep{newey1999two, ackerberg2014asymptotic}. Importantly, this fixed-dimensional result does not extend directly to the many weak interaction asymptotic regime in which $r$ diverges with $n$. This is because when the number of moments increases, estimation error in the nuisance parameters accumulates across moments. The next section develops estimation and inference procedures tailored to this challenging setting.

\section{Estimation under Many Weak Interaction Moments}
\label{sec:challenges}
\subsection{Continuous Updating Estimation with Known Nuisance}

Before introducing the feasible estimator, we briefly review the 
Continuous Updating Estimator (CUE) as a benchmark in the idealized 
setting where the nuisance parameters are known. This subsection 
serves two purposes. First, it clarifies the estimation framework under 
many interaction moment conditions. Second, it motivates our choice of 
CUE as the basis for the feasible MAGIC estimator developed below.

Modern genetic studies often involve a large number of candidate 
IVs, so that the number of interaction moment conditions $r$ 
may be substantial. Although these interactions may collectively explain 
variation in the exposure, individual interaction–exposure correlations 
may be weak \citep{wei2014detecting}. In such many weak moment settings, 
CUE 
\citep{newey2004higher} is particularly attractive because it jointly 
updates the parameter over the optimal weighting matrix, mitigating bias
under many weak moment asymptotics.

In the ideal case where the nuisance parameter is known, the CUE estimator is defined as
\[
\tilde{\beta}_{\mathrm{CUE}}
=
\argmin_{\beta \in \mathcal{B}}
\frac{1}{2}
\hat{m}(\beta)^\top
\hat{\Omega}^{-1}(\beta)
\hat{m}(\beta),
\]
where
\[
\hat{m}(\beta)=\mathbb{E}_n\{m(O;\beta,\mu^*)\},
\qquad
\hat{\Omega}(\beta)
=
\mathbb{E}_n\{m(O;\beta,\mu^*)^{\otimes 2}\}.
\]

In practice, however, the nuisance parameters 
are unknown and must be estimated. We develop a feasible 
plug-in version of CUE and address the 
additional challenges that arise when the number of interaction moments 
diverges with the sample size.

\subsection{Many Weak Moments and Nuisance Estimation}

Our setting introduces an additional complication: nuisance parameters must be estimated in a preliminary stage, and their estimation error propagates through a diverging number of weak interaction moments. This combination of many weak moment asymptotics and first-stage nuisance estimation differs fundamentally from both classical fixed-$r$ semiparametric theory and the many weak moment framework of \citet{Newey:2009aa}, which does not explicitly account for estimated nuisance parameters.

To illustrate the difficulty, consider a Taylor expansion of the first-order CUE condition around the true parameter value (see Section~\ref{sec:asymp} for details). The resulting expansion contains two components: a leading GMM term and a second-order degenerate U-statistic term.  For the GMM term, influence function-based asymptotic expansion  has to account for the additional complexity when the number of interactions $r$ grows along with $n$. 

In addition, while the second-order U-statistic term vanishes in probability under the classical setting with fixed $r$, it  becomes non-negligible under many weak moment asymptotics \citep{Newey:2009aa}. Under the latter regime, nuisance estimation error affects not only the empirical moment function but also its derivative with respect to the parameter of interest. Therefore further theoretical development is needed for valid inference, which includes controlling the biases of not only the empirical plug-in moment but also its derivative. This  motivates the stronger orthogonality requirement developed in the next subsection.

\subsection{Global Neyman Orthogonality}
\label{sec:global}

The difficulty identified in the previous subsection stems from the combination of two features: (i) a diverging number of weak moment conditions and (ii) first-stage estimation of nuisance parameters. Classical semiparametric theory controls nuisance-induced bias through Neyman orthogonality; see \citet{neyman1,neyman2} and modern treatments such as \citet{Chernozhukov2018ddml}. However, these results are developed under fixed-dimensional moment conditions and do not directly address the many weak moment regime considered here. To clarify the issue, consider first a generic linear moment function of the form
\[
\psi(O;\beta,\eta)
=
\psi^a(O;\eta)\beta
+
\psi^b(O;\eta),
\]
where $\eta$ denotes nuisance parameters. Under classical asymptotics with a fixed number of moments, Neyman orthogonality requires that
\begin{equation}
\label{eq:classical}
\mathbb{E}\!\left\{
\frac{\partial \psi(O;\beta^*,\eta^*)}{\partial \eta}
\right\}
=
0.
\end{equation}
This condition alleviates the biases from first-stage nuisance estimation for valid inference on the parameter of interest $\beta$.

When the number of moment conditions grows with the sample size, however, the situation changes fundamentally. In this regime, nuisance estimation error affects not only the empirical moment function but also its derivative with respect to $\beta$,
\[
\Psi(O;\eta)
=
\frac{\partial \psi(O;\beta,\eta)}{\partial \beta}
=
\psi^a(O;\eta).
\]

Both components enter the stochastic expansion of the CUE estimator. Controlling only \eqref{eq:classical} at $\beta=\beta^*$ is therefore insufficient: even if the moment function itself is locally orthogonal, the derivative may not be and contamination may persist.

To see this more explicitly, note that
\[
\mathbb{E}\!\left\{
\frac{\partial \psi(O;\beta,\eta^*)}{\partial \eta}
\right\}
=
\underbrace{
\mathbb{E}\!\left\{
\frac{\partial \psi^a(O;\eta^*)}{\partial \eta}
\right\}
}_{:=\Gamma}
\beta
+
\mathbb{E}\!\left\{
\frac{\partial \psi^b(O;\eta^*)}{\partial \eta}
\right\}.
\]
Classical orthogonality imposes that this expression vanishes at $\beta=\beta^*$, but it does not require that the coefficient $\Gamma$ itself be zero, so that the derivative may not be locally orthogonal. As the number of weak moment conditions increases, nuisance estimation error propagates through this derivative term, and the resulting bias can accumulate across moments. This accumulation can distort the stochastic expansion of the estimator and invalidate standard asymptotic approximations.

To eliminate this derivative contamination, we strengthen the orthogonality requirement and impose
\begin{equation}
\label{eq:global}
\mathbb{E}\!\left\{
\frac{\partial \psi(O;\beta,\eta^*)}{\partial \eta}
\right\}
=
0,
\quad \text{for all } \beta \in \mathcal{B}.
\end{equation}
We refer to \eqref{eq:global} as \emph{global Neyman orthogonality}. Unlike classical Neyman orthogonality \citep{neyman1,neyman2}, which holds only at the true parameter value, global Neyman orthogonality requires insensitivity of the moment function to nuisance perturbations uniformly over the parameter space. In particular, it implies $\Gamma=0$ and ensures that nuisance estimation error has asymptotically negligible impact on both the moment function and its derivative. Such uniform control becomes essential when the number of weak moments diverges.

\subsection{Orthogonalized Interaction Moments}

We now return to the interaction-based moment construction and show how an
appropriate orthogonalization yields the stronger property required in the
many weak interaction regime. Although the interaction-based  moments introduced
in Section~2 suffice for identification at the population level, they are not
well suited for estimation when the number of moment conditions diverges.
In this setting, nuisance estimation error can accumulate across moments and
affect both the moment function and its derivative with respect to the
parameter of interest. Consequently, the usual local Neyman orthogonality condition is generally insufficient.

To address this issue, we modify the interaction moments by projecting out
lower-order interaction components from both the outcome and the exposure
before forming higher-order interaction moments. This construction is related
to the partialling-out approach of \citet{robinson1988root} and to the
orthogonal score construction used in double machine learning
\citep{Chernozhukov2018ddml}. In contrast to those settings, which impose
orthogonality locally under fixed-dimensional moments, our setting involves a
diverging number of weak interaction moments and therefore requires a stronger
global orthogonality condition.

We implement this orthogonalization for each $k=2,...,p$ as follows. Define the lower-order non-demeaned interaction basis
\[
W_{k-1}(Z)
=
(1,\bar Z_{1}^\top,\dots,\bar Z_{k-1}^\top)^\top.
\]

Project the outcome and exposure onto $W_{k-1}(Z)$:
\[
f_k(Z;\theta_{k-1})=\theta_{k-1}^{\top}W_{k-1}(Z), 
\qquad
h_k(Z;\xi_{k-1})=\xi_{k-1}^{\top}W_{k-1}(Z),
\]
where $\theta_{k-1}$ and $\xi_{k-1}$ are the population coefficients from the
linear projections of $Y$ and $D$ onto $W_{k-1}(Z)$.
Define the population level residuals
\[
R_k^Y = Y - f_k(Z;\theta_{k-1}), 
\qquad
R_k^D = D - h_k(Z;\xi_{k-1}).
\]

The orthogonalized interaction moments are
\[
g(O;\beta,\eta)=(g_2^\top,\dots,g_q^\top)^\top,
\]
where
\begin{equation}
\label{eq:g}
g_k(O;\beta,\eta_k)
=
\bar{Z}_{k,\mu}
\big(
R_k^Y - \beta R_k^D
\big),
\quad k=2,\dots,q,
\end{equation}
and $\eta_k=(\mu,\theta_{k-1}^\top,\xi_{k-1}^\top)^\top$ collects the nuisance parameters. Thus the $k$th moment multiplies the centered interaction IV
$\bar Z_{k,\mu}$ with the residualized structural error $R_k^Y-\beta R_k^D$. 

Importantly, this orthogonalization does not alter identification: the $k$th  moment still captures the correlation of the $k$th order interactions with the exposure, as only interaction components of order lower than $k$  are partialled out. Identification is achieved as long as one such interaction correlates with the exposure, which is the same identifying condition
for $\beta^*$ based on original moment condition (\ref{eq:gmm}).

\begin{lemma}[Global Neyman Orthogonality]
Under Assumptions~\ref{assp:alice} and \ref{assp:indp},
\[
\mathbb{E}
\left\{
\frac{\partial g_k(O;\beta,\eta_k^*)}{\partial \eta_k}
\right\}
=
0,
\quad \text{for all } \beta \in \mathcal{B},
\]
for $k=2,\dots,q$.
\end{lemma}

Lemma~1 shows that our carefully constructed orthogonalized interaction moments satisfy global
Neyman orthogonality. The condition holds uniformly over $\beta\in\mathcal B$,
so nuisance estimation does not affect the moment function or its derivative
at first order even when the number of interaction moments diverges. This
property underlies the validity of the MAGIC estimator developed below.

\subsection{The MAGIC Estimator}

We now define the feasible estimator that incorporates first-stage nuisance estimation into the interaction-based CUE framework.

For each $k=2,\dots,q$, let
$
\hat{\eta}_k
=
(\hat{\mu},\hat{\theta}_{k-1}^\top,\hat{\xi}_{k-1}^\top)^\top
$
denote the estimator of $\eta_k^*$. Here $\hat{\mu}=\mathbb{E}_n(Z)^\top$ is the sample mean of the instruments, and $\hat{\theta}_{k-1}$ and $\hat{\xi}_{k-1}$ are ordinary least squares estimators obtained by regressing $Y$ and $D$, respectively, on $W_{k-1}(Z)$. These regressions implement the Robinson-style orthogonalization introduced in Section~4.3 by partialling out lower-order interaction components.

Although each nuisance estimator converges at the usual parametric rate, their estimation error propagates across the collection of interaction moments. When the number of interaction moments grows with the sample size, such accumulation must be controlled to ensure valid inference. As shown in Section~4.3, the interaction-based moment construction satisfies global Neyman orthogonality (Lemma~1), which ensures that first-stage estimation error does not affect the leading behavior of the objective function and its derivative, even in the many weak interaction regime. Formal growth conditions and asymptotic results are established in Section~5.

 The MAGIC estimator is the minimizer of the plug-in CUE objective:
\[
\hat{\beta}_{\mathrm{CUE}}
=
\argmin_{\beta \in \mathcal{B}}
\hat{Q}(\beta,\hat{\eta}),
\]
where
\[
\hat{Q}(\beta,\hat{\eta})
=
\frac{1}{2}
\hat{g}(\beta,\hat{\eta})^\top
\hat{\Omega}^{-1}(\beta,\hat{\eta})
\hat{g}(\beta,\hat{\eta}),
\]
\[
\hat{g}(\beta,\eta)
=
\mathbb{E}_n\{g(O;\beta,\eta)\},
\qquad
\hat{\Omega}(\beta,\eta)
=
\mathbb{E}_n\{g(O;\beta,\eta)^{\otimes 2}\}.
\]

Here,
$
\hat{\eta}
=
(\hat{\mu},\hat{\theta}_1^\top,\hat{\xi}_1^\top,\dots,
\hat{\theta}_{q-1}^\top,\hat{\xi}_{q-1}^\top)^\top
$
collects the first-stage nuisance estimators corresponding to
$
\eta^*
=
(\mu^*,\theta_1^{*\top},\xi_1^{*\top},\dots,
\theta_{q-1}^{*\top},\xi_{q-1}^{*\top})^\top.
$

\section{Theoretical Properties}
\label{sec:the}

This section develops the large-sample theory for the proposed MAGIC estimator. We study an asymptotic framework in which the number of interaction moments may grow with the sample size, so that identification is driven by the aggregate contribution of many potentially weak interaction–exposure associations. In this setting, nuisance estimation error can accumulate across moments, and standard fixed-dimensional GMM arguments no longer apply directly. We establish conditions for consistency and asymptotic normality, derive a consistent variance estimator, and show that the overidentifying restrictions test retains a chi-square limit under appropriate growth conditions.

\subsection{Many Weak Moment Asymptotics}
\label{sec:many}

We begin by specifying the asymptotic regime that governs the joint behavior of the number of moments and their strength. As the sample size increases, the dimension $r$ of the interaction moment vector may diverge, while individual interaction–exposure correlations may shrink. The following assumption formalizes this balance and ensures that the identification strength grows sufficiently fast for consistent estimation. Let $G(O;\eta)=\partial g(O;\beta,\eta)/\partial \beta$ denote the moment derivative and define
\[
\Omega(\beta,\eta)=\mathbb{E}\{ g(O;\beta,\eta)^{\otimes 2}\}, \quad 
\Omega=\Omega(\beta^\ast,\eta^\ast), \quad 
G(\eta)=\mathbb{E}\{G(O;\eta)\}, \quad 
G=G(\eta^\ast).
\]

\begin{assumption}
\label{assp:mwi}
There exists a sequence $\kappa_n\to\infty$ such that $r/\kappa_n^2$ is bounded and
\[
\frac{n G^\top \Omega^{-1} G}{\kappa_n^2}\rightarrow \lambda^2\in(0,\infty)
\quad \text{as } n\to\infty.
\]
\end{assumption}
Assumption \ref{assp:mwi} formalizes the idea that identification can arise from the combined contribution of many weak interaction moments. Even if individual interaction–exposure correlations are small, consistent estimation remains possible provided that their aggregate strength increases sufficiently fast at rate $\kappa_n$.

When $r$ is fixed, choosing $\kappa_n=\sqrt{n}$ reduces to conventional GMM asymptotics with strong moments. In contrast, when interaction moments are weak, identification depends jointly on the strength of interaction–exposure correlations and the number of interaction moments.

To illustrate, suppose $\Omega=I\sigma^2$, $G=c_1 n^{\varphi_1}\ell$, and $r=c_2 n^{\varphi_2}$ with $c_1,c_2>0$. Then
\[
n G^\top \Omega^{-1} G
= n^{1+2\varphi_1+\varphi_2}\left(\frac{c_1\sqrt{c_2}}{\sigma}\right)^2.
\]
If $\varphi_1=-1/2$ \citep{Staiger:1997aa} and $r$ is fixed with $\varphi_2=0$, identification fails because the effective strength encoded by $nG^\top \Omega^{-1} G$ does not diverge. However, when $r$ increases with $n$, Assumption~\ref{assp:mwi} can still hold. For example, if $r=c_2 n^{\varphi_2}$ and $\kappa_n^2=n^{\varphi_2}$, then $\lambda=c_1\sqrt{c_2}/\sigma$, illustrating a many weak IV regime \citep{chao2005consistent}.
Weaker interaction–exposure correlations (more negative $\varphi_1$) must therefore be offset by faster growth in the number of interaction moments (larger $\varphi_2$) to achieve the same rate $\kappa_n$.

More generally, Assumption~\ref{assp:mwi} holds with $\kappa_n^2=n^{1+2\varphi_1+\varphi_2}$ for parameter pairs $(\varphi_1,\varphi_2)$ in the shaded region of Figure~\ref{fig:rel}. The vertical axis indexes the decay rate of interaction–exposure correlations $\varphi_1$, while the horizontal axis indexes the growth rate of the number of interaction moments $\varphi_2$.  This highlights a key feature of the MAGIC framework: identification can be achieved through the aggregate contribution of many weak interaction moments rather than relying on any single strong interaction.

\begin{figure}[htbp] 
    \centering  
   \includegraphics[width=.5\textwidth]{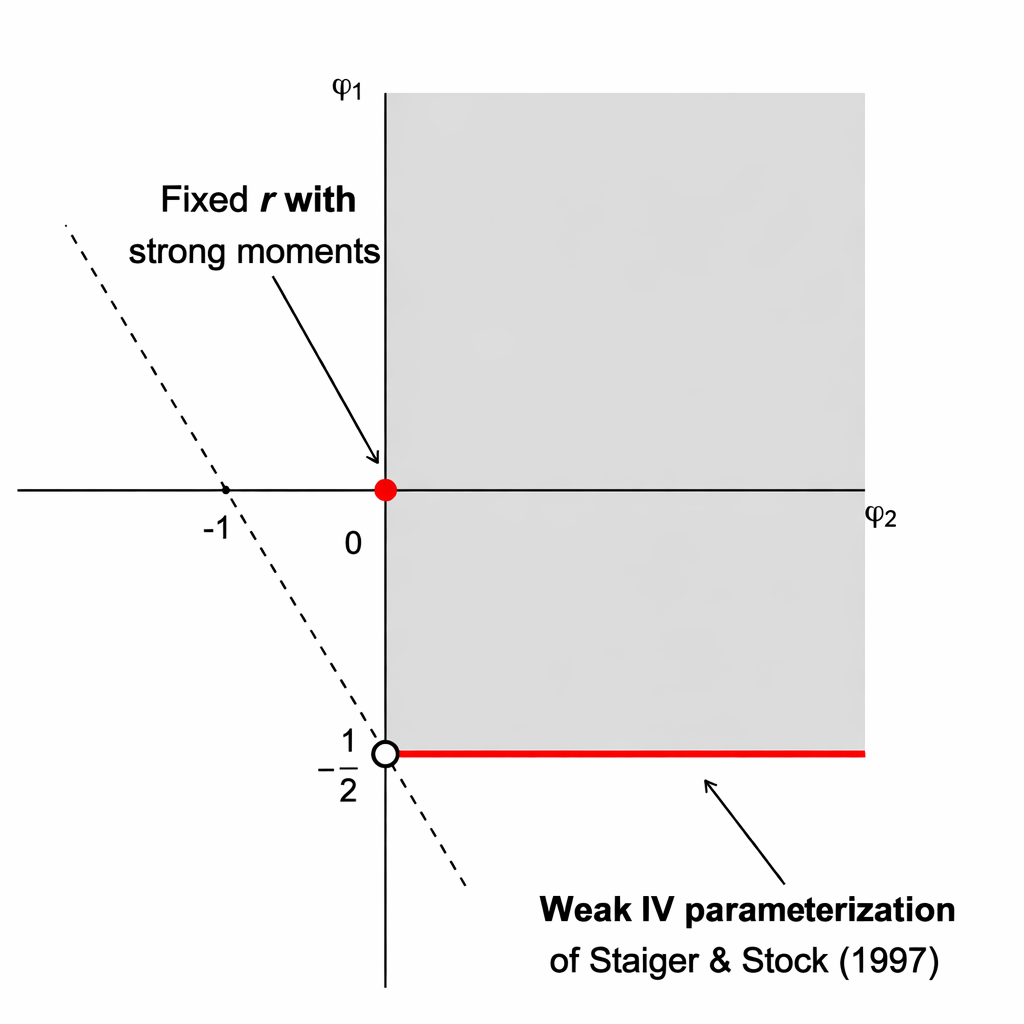} 
\caption{Illustration of the many weak interaction asymptotic regime. The vertical axis indexes the rate at which interaction–exposure correlations decay, while the horizontal axis indexes the growth rate of the number of interaction moments. The shaded region shows combinations of these rates for which the effective identification strength satisfy Assumption~\ref{assp:mwi}.}
    \label{fig:rel}  
\end{figure}

\subsection{Asymptotic Results}
\label{sec:asymp}

We now characterize the limiting distribution of the MAGIC estimator. The CUE estimator satisfies the first-order condition
\[
\kappa_n^{-2} n 
\frac{\partial \hat{Q}(\hat{\beta}_{\mathrm{CUE}}, \hat{\eta})}{\partial \beta}
= 0.
\]
A Taylor expansion around $\beta^*$ yields
\[
\kappa_n(\hat{\beta}_{\mathrm{CUE}} - \beta^*)
=
-
\left\{
n\kappa_n^{-2}
\frac{\partial^2 \hat{Q}(\bar{\beta}, \hat{\eta})}{\partial \beta^2}
\right\}^{-1}
\cdot
n\kappa_n^{-1}
\frac{\partial \hat{Q}(\beta^*, \hat{\eta})}{\partial \beta},
\]
where $\bar{\beta}$ lies between $\hat{\beta}_{\mathrm{CUE}}$ and $\beta^*$.

The asymptotic behavior of $\hat{\beta}_{\mathrm{CUE}}$ therefore depends on two components: the Hessian term and the gradient term. Suppose $\eta^*$ is known, then under regularity conditions, the Hessian satisfies
\[
n\kappa_n^{-2}
\frac{\partial^2 \hat{Q}(\bar{\beta},\eta^*)}{\partial \beta^2}
\xrightarrow{p}
\lambda^2 \in (0,\infty).
\]

The gradient term admits the decomposition
\[
n\kappa_n^{-1}
\frac{\partial \hat{Q}(\beta^*,\eta^*)}{\partial \beta}
=
\underbrace{
\sqrt{n}\kappa_n^{-1}
G^\top \Omega^{-1}
\sqrt{n}\mathbb{E}_n\{g(O;\beta^*,\eta^*)\}
}_{\text{GMM term}}
+
\underbrace{
\kappa_n^{-1}
\sum_{i\neq j}
\frac{U_i^\top \Omega^{-1} g(O_j;\beta^*,\eta^*)}{n}
}_{\text{degenerate U-statistic term}}
+ o_p(1),
\]
where $U_i$ is the population residual from least squares regression of the centered derivative $G_i-G$ on $g_i$, $i=1,2,...,n$. The first term corresponds to the familiar fixed-dimensional GMM component and contributes variance $\lambda^2$. The second term arises from the accumulation of many weak interaction moments and does not vanish asymptotically in this regime. It therefore introduces an additional variance component,
\[
\sigma^2
=
\lim_{n\to\infty}
\kappa_n^{-2}
\mathbb{E}(U^\top \Omega^{-1} U),
\]
which captures the extra variability induced by the many weak moment structure. 

When the nuisance parameter is estimated, global Neyman orthogonality ensures that nuisance estimation has asymptotically negligible impact on both components. Specifically,
\[
n\kappa_n^{-2}
\sup_{\beta\in\mathcal{B}}
\left|
\frac{\partial^2 \hat{Q}(\beta,\hat{\eta})}{\partial \beta^2}
-
\frac{\partial^2 \hat{Q}(\beta,\eta^*)}{\partial \beta^2}
\right|
=
o_p(1),
\]
and
\[
n\kappa_n^{-1}
\sup_{\beta\in\mathcal{B}}
\left|
\frac{\partial \hat{Q}(\beta,\hat{\eta})}{\partial \beta}
-
\frac{\partial \hat{Q}(\beta,\eta^*)}{\partial \beta}
\right|
=
o_p(1).
\]
Combining these results yields the following theorem.

\begin{theorem}[Consistency and Asymptotic Normality]
\label{thm:main}
Under Assumptions~\ref{assp:alice}--\ref{assp:mwi} and regularity conditions in the supplementary material, if $r^2/n \to 0$, then
$\hat{\beta}_{\mathrm{CUE}} \xrightarrow{p} \beta^*$.
If additionally $r^3/n \to 0$, then
\[
\kappa_n(\hat{\beta}_{\mathrm{CUE}} - \beta^*)
\;\xrightarrow{d}\;
N(0,V),
\qquad
V=\frac{\lambda^2+\sigma^2}{\lambda^4}.
\]
Moreover, the variance estimator $\hat V=\hat H^{-1}\hat D^\top \hat \Omega^{-1}\hat D \hat H^{-1}$ satisfies $\kappa_n^2 \hat V / n \xrightarrow{p} V$, where 
\[
\hat{H}
=
\frac{\partial^2 \hat{Q}(\hat{\beta}_{\mathrm{CUE}},\hat{\eta})}{\partial \beta^2},
\qquad
\hat{\Omega}
=
\hat{\Omega}(\hat{\beta}_{\mathrm{CUE}},\hat{\eta}),
\]
and
\[
\hat{D}
=
\mathbb{E}_n\{G(O;\hat{\eta})\}
-
\mathbb{E}_n\!\left\{
G(O;\hat{\eta})\,g(O;\hat{\beta}_{\mathrm{CUE}},\hat{\eta})^\top
\right\}
\hat{\Omega}^{-1}
\mathbb{E}_n\{g(O;\hat{\beta}_{\mathrm{CUE}},\hat{\eta})\}.
\]
\end{theorem}

\subsection{Test of Overidentifying Restrictions}

Our identification strategy relies on the interaction-based moment conditions implied by the ALICE structural model together with the independence assumption among the instruments. When $r \ge 2$, the model is overidentified, and the validity of these joint moment restrictions can be formally assessed.

In classical GMM theory, overidentifying restrictions are tested using a quadratic form of the empirical moments evaluated at the estimator \citep{sargan1958estimation, Hansen:1982aa}. In our setting, the natural analogue is the CUE objective function evaluated at $(\hat{\beta}_{\mathrm{CUE}},\hat{\eta})$. Under the null hypothesis
\[
H_0:\mathbb{E}\{g(O;\beta^*,\eta^*)\}=0,
\]
the population moment conditions hold, so the empirical objective should converge to zero at an appropriate rate.

Under many weak moment asymptotics, however, the limiting distribution of this quadratic form is not immediate. The number of moments $r$ may diverge with $n$, and nuisance estimation enters both the moment function and its derivative. It is therefore nontrivial that the classical chi-square limit continues to hold. The following theorem establishes that, under the same growth conditions required for asymptotic normality of the estimator, the overidentification test retains a standard chi-square distribution.

\begin{theorem}[Test of Overidentifying Restrictions]
\label{thm:overid}
Under the conditions of Theorem~\ref{thm:main} with $r^3/n \to 0$, and under
$
H_0:\mathbb{E}\{g(O;\beta^*,\eta^*)\}=0,
$
we have
\[
2n\hat{Q}(\hat{\beta}_{\mathrm{CUE}},\hat{\eta})
\;\xrightarrow{d}\;
\chi^2_{r-1}.
\]
An asymptotic level-$\alpha$ test rejects $H_0$ if
$
2n\hat{Q}(\hat{\beta}_{\mathrm{CUE}},\hat{\eta})
\ge
\chi^2_{r-1,\alpha},
$
where $\chi^2_{r-1,\alpha}$ denotes the $(1-\alpha)$ quantile of the $\chi^2_{r-1}$ distribution.
\end{theorem}
 Rejection of $H_0$ indicates that the interaction-based moment restrictions implied by the ALICE model are incompatible with the observed data.

\section{Simulation Studies}
\label{sec:sim}

We perform Monte Carlo experiments to investigate the finite-sample performance of the proposed MAGIC method under a study design which follows closely those  in \citet{Kang:2016aa}, \citet{Guo:2018aa}, \citet{Windmeijer:2019aa} and \citet{kang2020two}. Specifically, the putative IVs $Z_1,...,Z_{p}$ are generated from independent Bernoulli distributions with probability $\mu=0.5$, and $p=10$ or $20$ which corresponds to $r=45$ or $190$ distinct SNP two-way interactions, respectively. The outcome and exposure are then generated  from 
\begin{equation}
\label{eq:sim}
\begin{gathered}
    Y=D \beta +\overset{p}{\underset{j=1}{\sum}}\pi_j Z_j+\epsilon,\\
    D=\overset{p}{\underset{j=1}{\sum}}\theta_jZ_j+\overset{p-1}{\underset{k=1}{\sum}}\overset{p}{\underset{j=k+1}{\sum}}\alpha_{jk}Z_j Z_k+\nu,
\end{gathered}
\end{equation}
where $\beta=0$, $\alpha_{jk}=c n^{-1/2}$ with $c=3.75$, and $(\epsilon,\nu) \sim \mathcal{N}_2(0,\Sigma)$ with $\operatorname{Var}(\epsilon)=\operatorname{Var}(\nu)=1$ and $\operatorname{Cov}(\epsilon,\nu)=0.25$.
We implement the estimator $\hat{\beta}_{\text{CUE}}$ (MAGIC) with $q=2$ under the following scenarios:
\begin{itemize}
\item[I.]  (Majority \cmark,  InSIDE \xmark) $\theta_j=1$ for $j=1,...,p$, and $30\%$ of the SNPs are invalid, with $\pi_j=0.2$.
\item[II.]  (Plurality \cmark,  InSIDE \xmark) $\theta_j=1$ for $j=1,...,p$, and $60\%$ of the SNPs are invalid: $20\%$ with $\pi_j=0.2$, $20\%$ with $\pi_j=0.4$ and the remaining $20\%$ with $\pi_j=0.6$.
\item[III.] (Plurality  \xmark, InSIDE \cmark) $\theta_j$ are independently sampled from $N(1,1)$, and $\pi_j$ are  independently sampled from $N(0.2,0.2)$, for $j=1,...,p$.
\item[IV.] (Plurality  \xmark, InSIDE \xmark) $\theta_j$ are independently sampled from $N(1,1)$ for $j=1,...,p$, and $70\%$ of the SNPs are invalid with $\pi_j=\theta_j/2$.
\end{itemize}
For comparison, we also implement the two-stage least squares ({TSLS}) estimator \citep{angrist1995two}  and the following competing methods under default settings: Two-Stage Hard Thresholding (TSHT) \citep{Guo:2018aa}, and adaptive Lasso (A-Lasso) \citep{Windmeijer:2019aa}.

The Monte Carlo results based on 1000 replications are summarized in Table \ref{tab:1} for $n=5000$. In agreement with theory, A-Lasso and TSHT perform well in terms of bias and coverage when the majority and plurality rules hold, respectively. The proposed MAGIC method shows  small bias and  coverage close to nominal levels across the scenarios. The supplementary material contains additional Monte Carlo results for sample size $n=20000$ with qualitatively similar conclusions. 

\begin{table}[htbp]
\renewcommand{\arraystretch}{0.9}
\centering
\caption{Monte Carlo results for estimation of $\beta$, $n=5000$.}
\label{tab:1}
\resizebox{1\columnwidth}{!}{
\begin{tabular}{ll rrrr rrrr}
\hline
 &  & \multicolumn{4}{c}{$p=10$} & \multicolumn{4}{c}{$p=20$} \\
\cline{3-6}\cline{7-10}
Scenario & Method 
& $|\text{Bias}|$ & SD & Mean SE & Coverage (95\%) 
& $|\text{Bias}|$ & SD & Mean SE & Coverage (95\%)\\
\hline

\multirow{4}{*}{I}
& MAGIC & .006 & .336 & .331 & .966 & .009 & .253 & .185 & .958 \\
& TSLS     & .060 & .005 & .004 & .000 & .060 & .006 & .006 & .000 \\
& TSHT     & .000 & .005 & .005 & .952 & .003 & .010 & .007 & .843 \\
& A-Lasso  & .000 & .005 & .005 & .953 & .002 & .008 & .008 & .926 \\
\hline

\multirow{4}{*}{II}
& MAGIC & .008 & .410 & .364 & .936 & .003 & .154 & .139 & .960 \\
& TSLS     & .240 & .005 & .005 & .000 & .240 & .007 & .007 & .000 \\
& TSHT     & .000 & .007 & .007 & .934 & .026 & .031 & .018 & .677 \\
& A-Lasso  & .200 & .011 & .010 & .000 & .199 & .014 & .014 & .000 \\
\hline

\multirow{4}{*}{III}
& MAGIC & .007 & .298 & .285 & .948 & .007 & .132 & .134 & .950 \\
& TSLS     & .105 & .053 & .004 & .011 & .103 & .038 & .005 & .004 \\
& TSHT     & .133 & .226 & .029 & .135 & .134 & .163 & .043 & .208 \\
& A-Lasso  & .131 & .098 & .007 & .045 & .127 & .064 & .009 & .023 \\
\hline

\multirow{4}{*}{IV}
& MAGIC & .024 & .645 & .510 & .960 & .006 & .266 & .171 & .959 \\
& TSLS     & .352 & .084 & .012 & .000 & .352 & .062 & .005 & .000 \\
& TSHT     & .489 & .050 & .013 & .000 & .499 & .015 & .006 & .000 \\
& A-Lasso  & .496 & .030 & .014 & .004 & .500 & .007 & .006 & .000 \\
\hline

\end{tabular}}
\vspace{0.2cm}

\begin{tablenotes}
\footnotesize
\item \textit{Notes:} Scenario I assumes majority validity with InSIDE violated; 
Scenario II assumes plurality validity with InSIDE violated; 
Scenario III assumes plurality invalidity with InSIDE satisfied; 
Scenario IV assumes plurality invalidity with InSIDE violated. 
$|\text{Bias}|$ and SD denote the Monte Carlo absolute bias and standard deviation, respectively.
Mean SE is the mean estimated standard error, and Coverage (95\%) is the empirical coverage of the 95\% Wald confidence intervals, based on 1000 replications. 
Entries reported as 0 indicate values smaller than 0.0005. 
\end{tablenotes}
    
\end{table}

\begin{table}[!htbp]
\renewcommand{\arraystretch}{0.9}

\centering
\caption{Monte Carlo results for measure of weak identification and diagnosis.}  
    \label{tab:3}
\resizebox{.8\columnwidth}{!}{
\begin{tabular}{ p{6em}r  rrrr  rrrrr }
 \hline
   \noalign{\smallskip}
 Scenario & $c$ &$n$ &  $\hat{F}_2$& $|\text{Bias}|$ & SD & Mean SE & Coverage (95\%) & Empirical size \\ 
 \hline
 \multirow[c]{10}{*}{\shortstack{ALICE\\ \phantom{\cmark}model \cmark}} &  \multirow[c]{2}{*}{\shortstack{0.00}}  &  5000& 1.030& .296& 2.326&  4.628& .940&  .014\\ 
 & &  20000& .995&  .335& 2.368&  4.298& .949& .013\\ 

   \noalign{\smallskip}
&\multirow[c]{2}{*}{\shortstack{2.50}}  
&  5000& 1.441&  .063& 1.025&  1.053& .961& .005\\ 
&&  20000& 1.389&  .053& 1.028&  1.162& .961& .017\\ 
   \noalign{\smallskip}
 &\multirow[c]{2}{*}{\shortstack{3.75}}  
 &  5000& 1.928&  .050& .428&  .400& .968& .027\\ 
 & &  20000& 1.897&  .024& .333&  .301& .964& .030\\ 
    \noalign{\smallskip}
&  \multirow[c]{2}{*}{\shortstack{5.00}} 
&  5000& 2.630&  .009& .165& .168& .959& .026\\ 
 & &  20000& 2.591&  .006& .161& .160& .962&.030\\ 
    \noalign{\smallskip}
&  \multirow[c]{2}{*}{\shortstack{7.50}} 
&  5000& 4.632&  .005& .091& .093& .962& .048\\ 
&  &  20000& 4.555& .005& .093& .091& .946&.032\\

\hline

\multirow[c]{10}{*}{\shortstack{ALICE\\ \phantom{\xmark}model \xmark}} &\multirow[c]{2}{*}{\shortstack{2.50}}  
&  5000& 1.438&  2.041& 1.811&  2.004& .364& .128\\ 
&&  20000& 1.403& 2.144& 1.495&  1.626& .358& .131\\ 
   \noalign{\smallskip}
 &\multirow[c]{2}{*}{\shortstack{3.75}}  
 &  5000& 1.935&  2.147& .725&  .683& .968& .334\\ 
 & &  20000& 1.905&  2.087& .624& .601& .964& .314\\ 
    \noalign{\smallskip}
&  \multirow[c]{2}{*}{\shortstack{5.00}} 
&  5000& 2.631&  2.069& .450& .424& .002& .620\\ 
 & &  20000& 2.582&  2.058& .442& .394& .000&.669\\ 
    \noalign{\smallskip}
&  \multirow[c]{2}{*}{\shortstack{7.50}} 
&  5000& 4.622& 2.045& .278& .277& .000& .981\\ 
&  &  20000& 4.567& 2.002& .234& .232& .000&.993\\ 
\hline
 \end{tabular}}

 \begin{tablenotes}
      \item  {\noindent\footnotesize Note: $\hat{F}_2$ is the average heteroscedasticity-robust $F$-statistic, Empirical size is the rejection rate of the overidentification test at level $\alpha=0.05$, based on 1000 repeated simulations. The definitions of other parameters are explained in the footnote of Table \ref{tab:1}. Zeros denote values smaller than $.0005$. 

}
    \end{tablenotes}
    
\end{table}

{
We investigate the effect of identification strength on the performance of MAGIC in a separate set of Monte Carlo experiments with $p=10$. Specifically, the outcome and exposure are generated from linear models (\ref{eq:sim}), with $\theta_j$ independently sampled from $N(1,1)$, and $\pi_j$ independently sampled from $N(0,0.2)$, for $j=1,...,p$. We vary the value of $c$ and compute  the heteroscedasticity-robust $F$-statistic  $\widehat{F}_q$ in the regression of $\bar{D}$, which is the residual after partialling out the linear effect of ${Z}$, on $(1,\bar{Z}^\top _{2,\hat{\mu}},...,\bar{Z}^\top _{q,\hat{\mu}})^\top$ evaluated at sample means. It serves as an informal, empirical measure of the exposure variation due to the  interactions of candidate IVs, up to the $q$-th order.

We also evaluate the proposed test for overidentifying restrictions with asymptotic type I error $\alpha=0.05$, based on the test statistic $2n\hat{Q}(\hat{\beta}_{\text{CUE}},\hat{{\eta}})$. For violation of the ALICE model, we consider:
$$Y=D \beta +\overset{p}{\underset{j=1}{\sum}}\pi_j Z_j+\underset{k=1}{\overset{p-1}{\sum}}\underset{j=k+1}{\overset{p}{\sum}}\phi_{jk} {Z}_j Z_k +\epsilon,$$
where $\phi_{jk}=\varepsilon_{jk}{c}{n}^{-1/2}$ and $\varepsilon_{jk}$'s are constants generated from $N(1,1)$.
The results based on 1000 replications are summarized in Table~\ref{tab:3}. The bias of MAGIC decreases as the heteroskedasticity-robust first-stage statistic $\widehat{F}_{q}$ increases for $q=2$, consistent with stronger interaction–exposure relevance and improved identification strength. When the ALICE model is correctly specified, the empirical rejection rate of the overidentification test remains close to the nominal level, indicating appropriate size control under many weak moment asymptotics. When the ALICE model is misspecified, the rejection frequency increases with the magnitude of the violation parameter $c$, demonstrating that the test has increasing power to detect structural departures from the assumed model.

\section{Application to UK Biobank Data}
\label{sec:app}

In this section, we employ MAGIC to estimate the causal effects of BMI on blood glucose level in the UK Biobank cohort, which is a large-scale biomedical database that includes in-depth genetic and health information from around 500,000 participants across the UK \citep{bycroft2018uk}. In our analysis, we included individuals aged 50 years or older who (1) did not report using antihypertensive medication and (2) had no missing data, resulting in a  sample of $n=197,144$ participants. We selected the top 5 significant and independent single nucleotide polymorphisms (SNPs) associated with BMI, after applying linkage disequilibrium (LD) clumping ($r^2<0.01$), as candidate IVs. For consistency with the simulation study, we included the same competing methods, namely two-stage least squares (TSLS), two-stage hard thresholding (TSHT), and adaptive Lasso (A-Lasso).

The results are summarized in Table~\ref{tab:app}. In this application, we set $q=2$, yielding $r=10$ pairwise genetic interaction terms. The corresponding heteroskedasticity-robust first-stage F-statistic is $\widehat{F}_2 = 1.227$, indicating limited interaction–exposure strength in this dataset. MAGIC yields a positive point estimate of $0.272$ for the effect of BMI on blood glucose ($p$-value $= 9.03\times 10^{-3}$). This estimate is larger and associated with a larger standard error than those obtained from alternative MR procedures based on main SNP effects. The difference reflects the distinct source of identifying variation in MAGIC. Whereas conventional estimators rely on main genetic effects and may be sensitive to violations of exclusion or independence assumptions, MAGIC exploits higher-order interaction moments that are robust to such violations under the ALICE model. If the main effects are partially contaminated by such violations, interaction-based moments may recover the true causal effect. At the same time, the interaction–exposure correlations are not strong, as indicated by the first-stage diagnostics, so identification relies on many weak moments, which increases sampling variability. This pattern illustrates the efficiency–robustness trade-off inherent in assumption-lean identification strategies: robustness to invalid instruments may come at the cost of reduced efficiency when identifying signals are weak. The $p$-value for the overidentifying restrictions test is $0.821$, providing no evidence against the maintained moment conditions.

\begin{table}[H]
\renewcommand{\arraystretch}{0.9}
 
\begin{center}
\caption{Estimated effects of BMI on blood glucose level in UK Biobank.}  
    \label{tab:app}
\begin{tabular}{ cccc}
 \hline
  \noalign{\smallskip}
MAGIC & TSLS & TSHT & A-Lasso \\
 \hline 
 .272 $\pm$ .205 &  .056 $\pm$ .018 & .056 $\pm$ .019 & .056 $\pm$ .018\\
\hline
\end{tabular}
\end{center}
\begin{tablenotes}
      \item  {\noindent\footnotesize Note: Point estimate $\pm$ 1.96$\times$standard error.

}
\end{tablenotes}
\end{table}

\section{Discussion}
\label{sec:discussion}

This paper develops a constructive approach to IV identification and inference, motivated by challenges arising in Mendelian randomization. By exploiting interaction-derived moment conditions among independent IVs, MAGIC enables identification without requiring any of the IVs to be valid. In contrast to existing approaches that rely on structural assumptions about IV validity \citep{kang2024identification}, identification in our framework follows from empirically verifiable independence and interaction construction, and is therefore assumption-lean in the sense that it avoids restrictions on the number or distribution of valid IVs.

On the theoretical side, we extend semiparametric GMM theory \citep{Hansen:1982aa,han2006gmm,Newey:2009aa,ye2024genius} to settings with many weak moment conditions and a diverging dimension of estimated nuisance parameters. We introduce a global Neyman orthogonality condition that ensures nuisance estimation has asymptotically negligible impact on both the moment function and its derivative, a requirement that becomes essential when the number of moment conditions grows with the sample size. By formalizing and controlling derivative bias in this regime, our framework advances the theory of weak identification and contributes to the broader literature on  semiparametric inference.         

Several directions merit further investigation. Relaxing the independence assumption among IVs, for example by allowing weak or structured forms of dependence, is of practical interest and would broaden the applicability of the constructive identification strategy. In addition, while we characterize the semiparametric efficiency bound under fixed-dimensional settings, deriving an efficiency bound under many weak moment asymptotics remains an open problem. Addressing this question would further deepen the theoretical understanding of semiparametric inference in settings with many weak moment conditions.

\section*{Acknowledgement}
This research has been conducted using the UK Biobank Resource under application number 52008. Zhonghua Liu was supported by NIH grant under award R01 AG086379. Baoluo Sun was supported by Singapore MOE AcRF Tier 1 grant A-8002935-00-00.

  \bibliography{refs.bib}

\end{document}